\documentclass[sort&compress,3p]{elsarticle}

\usepackage{graphicx}
\usepackage{amsmath}
\usepackage{amssymb}

\def\be{\begin{equation}}
\def\ee{\end{equation}}     
\def\bfi{\begin{figure}}
\def\efi{\end{figure}}
\def\bea{\begin{eqnarray}}
\def\eea{\end{eqnarray}}

\usepackage{hyperref}

\begin{document}

\title{Condensation and equilibration in an urn model}

\author[sa]{Federico Corberi\corref{cor}}
\ead{corberi@sa.infn.it}

\author[ba]{Giuseppe Gonnella}

\author[ba]{Alessandro Mossa}

\cortext[cor]{Corresponding author}
\address[sa]{Dipartimento di Fisica ``E.~R.~Caianiello'', and INFN, Gruppo Collegato di Salerno, and CNISM, Unit\`a di Salerno,
Universit\`a  di Salerno, via Giovanni Paolo II 132, 84084 Fisciano (SA), Italy.}
\address[ba]{Dipartimento di Fisica, Universit\`a di Bari and
INFN, Sezione di Bari, via Amendola 173, 70126 Bari, Italy.}

\begin{abstract}
After reviewing the general scaling properties of aging systems,
we present a numerical study of the slow evolution induced in the
zeta urn model by a quench from a high temperature to a lower one where
a condensed equilibrium phase exists. By considering both one-time
and two-time quantities we show that the features of
the model fit into the general framework of aging systems.
In particular, its behavior can be interpreted in terms of the simultaneous
existence of equilibrated and aging degrees with different scaling properties.
\end{abstract}


\maketitle

\section{Introduction}\label{intro}

Slow relaxation \cite{slow1,slow2,slow3,slow4,slow5,slow6} is the process whereby a system evolves without
reaching equilibrium in finite times. 
Physical realizations of this phenomenon
are observed in many materials, for instance ferromagnets, binary or complex fluids 
and glasses, when they are subjected to a drastic
change of thermodynamic control parameters, as in a
temperature quench or a pressure crunch. 
In the large-time regime these systems exhibit typical features of equilibrium, such as the 
absence of macroscopic currents and the constancy of one-time observables, together with 
hallmarks of non-stationarity, like the lack of time-reversal invariance and the 
dependence of two-time quantities on the system's age $t_w$, i.e., the time elapsed after 
the initial change of parameters.

These apparently contrasting features can be interpreted in a coherent
way if the degrees of freedom of the non-equilibrium state can
be classified into groups with different properties \cite{MVZ,FV}. In the simplest
cases, including ferromagnetic systems and the model considered in this
paper, this amounts to identify those degrees whose evolution is 
sufficiently fast and unrelated to the global non-equilibrium
condition of the system to rapidly self-equilibrate with
respect to the new value of the control parameters. These \emph{fast} degrees,
whose collection we will indicate with $\{ \psi \}$ in the following, are
responsible for the appearance of the typical features of equilibrated systems
discussed above. The rest of the system -- constrained degrees which are
not able to efficiently equilibrate with respect to the new control
parameters -- are a \emph{slow} component, denoted here with $\{ \sigma \}$, 
which gives rise to the aging properties, namely non-stationarity and
lack of time-reversal invariance. 

This separation into two classes is fundamental to understand
the properties of aging systems: the possibility to disentangle global
properties into their  very different contributions
has represented a groundbreaking tool
towards the recognition of the ultimate scaling structure
underlying the evolution. However, despite the relevance of the subject,
a precise operative definition of fast and slow degrees is, in general, difficult
to adopt and a convincing proof of their existence has been given, as to now,
only in few cases \cite{noi2002}.

In this paper we consider a simple paradigmatic example of aging dynamics, the zeta urn model \cite{bialas1,bialas2,bialas3},
where fast and slow degrees can be easily identified, their different properties
can be studied, and a clear picture of their mutual interplay can be drawn.
The scaling properties of $\{ \psi \}$ and $\{ \sigma \}$ are explicitly worked out
and the relative contribution to the behavior of different observables, either 
dependent on a single time or on two times, is discussed. We also discuss how
different and \emph{a priori} unrelated operative definitions of $\{\psi\}$ and $\{\sigma\}$
lead to the same overall properties, confirming  that the separation of degrees is a robust
and general property of aging which does not depend on the details of the procedure.

The paper is organized as follows:
In section \ref{twotime} we review on general grounds the behavior of one-time and 
two-time quantities, showing how they differently encode the non-equilibrium
properties. In section \ref{statag} a general identification of slow and fast 
components is made, their statistical properties are worked out, and
their mirroring in physical observables is analyzed.
In section \ref{urnmod} we introduce the zeta urn model that will be considered
in the remaining of the paper. The result of a numerical study of the model is
presented in the following section \ref{numerical}, where we explicitly show
a consistent way to identify the sets $\{ \psi \}$ and $\{ \sigma \}$ and explore
their contribution to one-time and two-time quantities.
Section \ref{concl} contains our conclusions and the discussion of some
open issues. 

\section{One-time and two-time quantities} \label{twotime}

Let us denote with $n=n_\psi (t)+n_\sigma (t)$  the number of degrees of freedom 
which are separated into the fast set $\{\psi\}$ and the slow one $\{\sigma\}$.
Since the system is approaching
equilibrium, where by definition out-of-equilibrium modes are
absent, the \emph{measure} of $\{ \sigma \}$ 
is bound to decrease as the evolution proceeds, 
\be
\lim _{t\to \infty}n_\sigma (t)=0.
\label{nsig0}
\ee 
This occurs independently of the capacity of the system to reach equilibrium in 
a finite time (interrupted aging) or not. In the following we will always consider
the second case, where equilibration is never achieved, which requires the thermodynamic
limit $n\to \infty$ to be taken before the large-time limit $t\to \infty$.

The organization of the internal modes into two sets
is reflected in the behavior
of any physical observable ${\cal O}$. In the case in which 
$\{ \psi \}$ and $\{ \sigma \}$ can be considered statistically independent,
an additive form 
\be
{\cal O}={\cal O}_\psi + {\cal O}_\sigma
\label{additive}
\ee
is found, with obvious notation. 

Let us consider one-time quantities first,
as for instance the internal energy, the pressure  or the density.
Due to the equilibrated character of $\{ \psi \}$,
${\cal O}_\psi$ reaches almost immediately its time-independent
equilibrium value ${\cal O}_{eq}$.
For example, taking ${\cal O}(t)$ as the energy,
${\cal O}_{eq}=E_{eq}$ represents the internal energy that the system would possess in the final 
equilibrium state. 
If ${\cal O}$ is sufficiently {\it regular}, in a sense that will be more precise at the end of
this section, since $n_\sigma (t)\to 0$ 
for sufficiently long times, 
one generally has 
\be 
\lim _{t\to \infty} {\cal O}_\sigma (t)=0.
\label{eqquant}
\ee
For this reason, if $t$ is large enough
and ${\cal O}_{eq}$ is finite, the contribution of
${\cal O}_\sigma$ represents only a small vanishing correction 
and this makes one-time observables basically indistinguishable from their 
constant equilibrium value. This however does not mean that
the system is equilibrated, but only that the contribution of 
the $\sigma$-s is negligible in the one-time quantities.
Clearly, if ${\cal O}_{eq}$ is known the small aging contribution 
${\cal O}_\sigma$ can be evidenced by subtracting
${\cal O}_{eq}$ from ${\cal O}$. 

The non-equilibrium aging character is instead more manifest in the
behavior of two-time quantities (again sufficiently {\it regular}, see below)
${\cal O}(t,t_w)$, with $t\ge t_w$, such as correlation functions or responses.
In the asymptotic time-domain in which 
the age $t_w$ of the sample
is large, the typical timescales $t_\psi$ and $t_\sigma$ 
over which fast and slow degrees decorrelate are widely separated. 
Indeed, while $t_\sigma=t_\sigma (t_w)$ is usually an increasing function of $t_w$,
$t_\psi$ is generally microscopic and 
age-independent, again due to the fact that $\{\psi \}$ is basically
in equilibrium.  

\begin{figure}
    \centering
\rotatebox{0}{\resizebox{.45\textwidth}{!}{\includegraphics{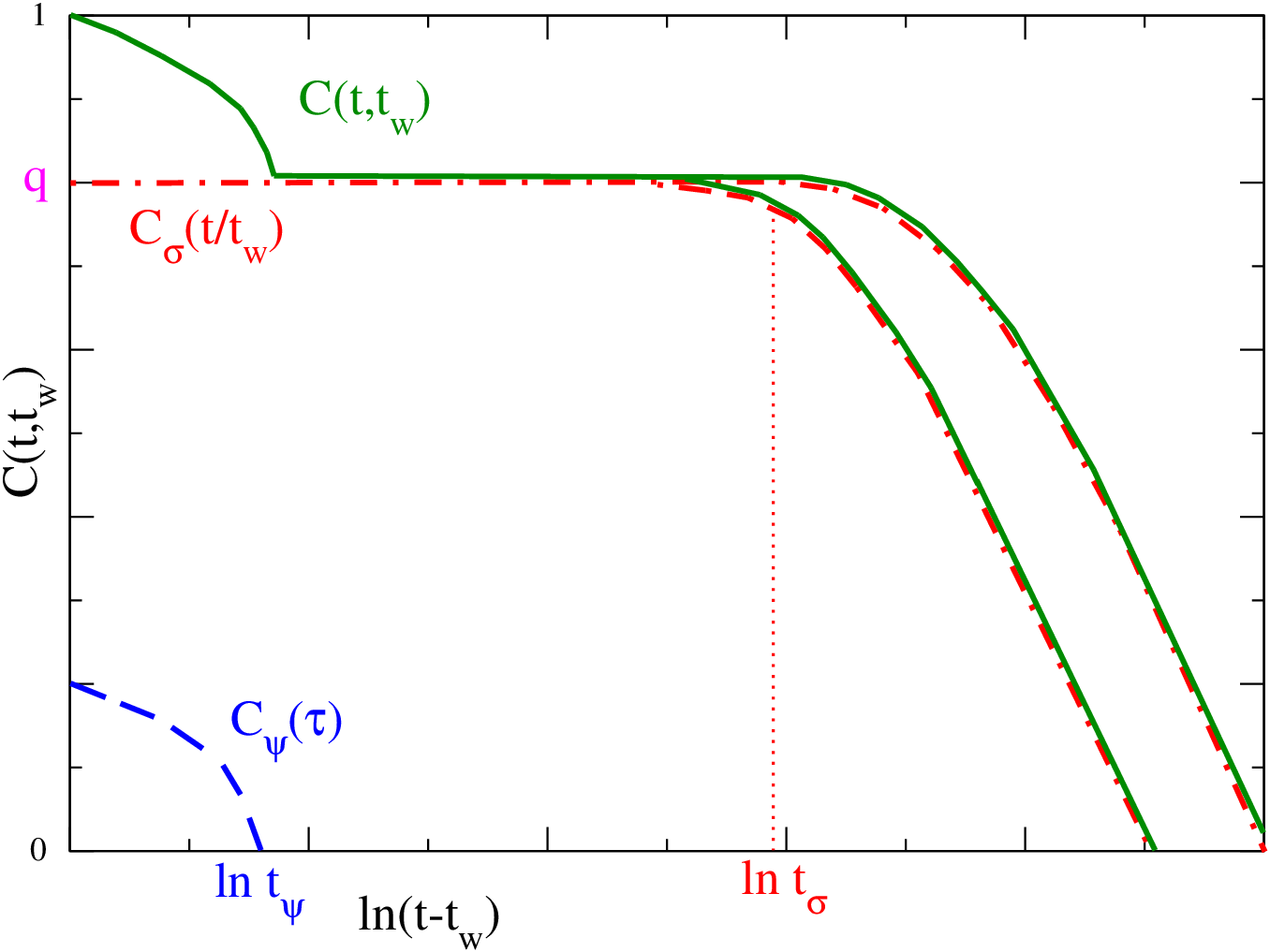}}}
\rotatebox{0}{\resizebox{.46\textwidth}{!}{\includegraphics{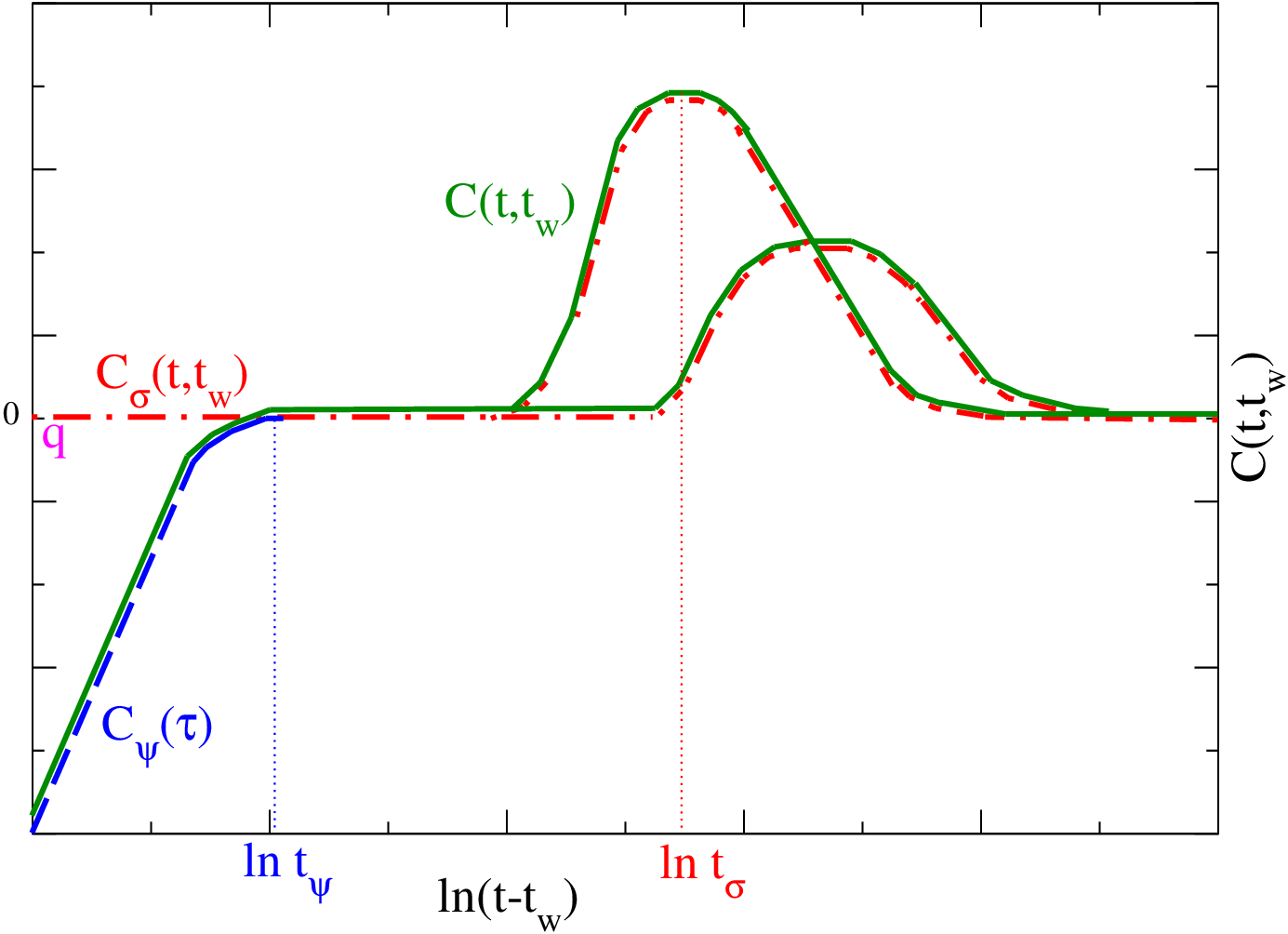}}}
\caption{The behavior of $C_\psi(\tau)$, $C_\sigma (t,t_w)$ and $C(t,t_w)$,
as $\tau =t-t_w$ is varied, is schematically shown on a log-log plot.
The left panel is for a ferromagnetic system, the right one for the urn model. 
For each function two
waiting times are considered, the curves corresponding to the larger
$t_w$ decay later (the curves for $C_\psi$ collapse on a single one because
this quantity is stationary). Note that, as discussed in the text, for a ferromagnet
$C_\sigma(t,t_w)$ only depends on $t/t_w$.}
\label{figcorr}
\end{figure}

This is made evident by the fact that a two-time quantity ${\cal O}(t,t_w)$ can probe 
either fast or slow degrees 
according to the direction along which the large-time limit is approached 
in the $(t,t_w)$ plane.
Specifically, if we let $t_w \to \infty$ while keeping
$\tau =t-t_w$ finite, since $\tau \ll t_\sigma$ 
the slow degrees act adiabatically 
and can be considered as static. In this time sector,
denoted as the short time difference (or quasi-equilibrium) regime, 
one sheds light on the dynamical behavior of fast modes.
Indeed ${\cal O}_\psi (\tau)$, which depends only on $\tau =t-t_w$
due to the stationarity of $\psi$, converges
to its asymptotic equilibrium value $\lim _{\tau \to \infty}{\cal O}_\psi=
{\cal O}_{eq}$
in a rather short time $\tau \simeq t_\psi$ in which ${\cal O}_\sigma $ 
cannot change appreciably and remains fixed to its equal-time value ${\cal O}_\sigma (t_w,t_w)=q$.
This is pictorially shown in the left panel of Fig.~\ref{figcorr}, where the behavior of a typical two-time 
quantity, the autocorrelation function $C(t,t_w)$ of a spin system (see the definition in Eq.~(\ref{autoco})),
and its contributions $C_\psi(t,t_w),C_\sigma (t,t_w)$ are sketched. The short time difference sector
is on the left part of the picture.

Conversely, letting again $t_w \to \infty$ but keeping $t/t_w$ 
(or the ratio $\ell(t)/\ell(t_w)$, where $\ell$ is a monotonously increasing function, 
in specific systems)
finite, one has $\tau \sim  t_w \sim t_\sigma$: This means that the analysis is focused on the 
timescale of the slow (or aging) components. In this time
sector, usually denoted as aging regime (on the right side of the left panel of of Fig. \ref{figcorr}), the fast degrees act as a stationary 
background and the time evolution of the slow modes is probed. 
Indeed, from $\tau \simeq t_\sigma $ onward
also ${\cal O}_\sigma $ starts changing and ${\cal O}$ moves from the
plateau ${\cal O}(t,t_w)\simeq q$ to the final 
value $\lim _{\tau \to \infty}{\cal O}(t,t_w)$.

Generally, a good separation of the degrees can only be operated for very large times.
The previous discussion shows that, in this case, two-time observables are quantities particularly fitted 
to highlight the role played by $\psi $ and $\sigma $.

We mentioned before that $\lim _{t\to \infty} {\cal O}_\sigma =0$ occurs for 
sufficiently {\it regular } observables ${\cal O}$. We now precise this point.
Consider, in order to keep the discussion simple, a quantity that
can be expressed as ${\cal O}=\sum _i o^{(i)}$, where the index $i$ runs over the
microscopic degrees of freedom and $o^{(i)}$ is the contribution of one such
degree. This is usually the case when dealing with extensive quantities.
Then we can write ${\cal O}_\sigma (t)=\sum _{i\in \{\sigma \}}o_\sigma ^{(i)}(t)=
n_\sigma (t)\overline o_\sigma (t)$, where $\overline o_\sigma (t)$
is the {\it average} contribution to ${\cal O}_\sigma $ provided by one slow degree.
If the time-dependence of $\overline o_\sigma (t)$ is such that
$\lim _{t\to \infty}n_\sigma (t)\overline o_\sigma (t)=0$,
then $\lim _{t\to \infty}{\cal O}(t)=\lim _{t\to \infty}{\cal O}_\psi (t)={\cal O}_{eq}$
and we call ${\cal O}$ an {\it equilibrating } quantity.
Otherwise, the contribution of ${\cal O}_\sigma $ does not disappear and 
$\lim _{t\to \infty}{\cal O}\neq {\cal O}_{eq}$.

Depending on the system considered and on the nature of ${\cal O}$ 
either situation can be observed. In most cases $\overline o$ is weakly dependent on time,
and the corresponding observable is an equilibrating one. For instance the energy of 
a non-disordered magnet quenched to a temperature $T_f$ below the critical temperature converges 
to the value $E_{eq}(T_f)$ provided by the standard equilibrium statistical mechanical 
calculation. However there are also some notable exceptions, as for instance the 
response function $\chi (t)$ (a quantity deeply related to the magnetic energy)
of a disordered one-dimensional magnet quenched to $T_f=0$, 
as described by the Ising chain with a random
magnetic field of strength $h$.
In the limit $h\to 0$ one finds~\cite{noirf1,noirf2,noirf3,noirf4,noirf5} 
$\lim _{t\to \infty}\chi(t)\neq \chi_{eq}$.

\section{Identification of stationary and aging modes} \label{statag}

A possible identification of fast and slow degrees, based on arguments
similar to the ones discussed in the previous section, can be 
given along the lines of Ref.~\cite{FV}. 
Referring to a system described by a collection
$S_i=\pm 1$ of spin variables one defines the slow part of $S_i$
in terms of the running time-average
\be\sigma _i(t) \simeq \frac{1}{\Delta t}\int _t ^{t+\Delta t} dz\, S_i(z)
\label{runav}
\ee
where 
\be
t_\psi \ll \Delta t \ll t_\sigma .
\label{condrunav}
\ee
The fast part is then obtained by subtraction
\be
\psi _i \simeq S_i(t)-\sigma _i(t).
\label{defpsi}
\ee
The idea in Eq.~(\ref{runav}) is that, since
$\Delta t$ is chosen much larger than $t_\psi$ the contribution to $S_i$
of the fast modes self-averages to zero (in a more general situation the fast degrees may average 
to a finite constant value) and we are left precisely
with the contribution of the slow modes.

The condition $\Delta t\ll t_\sigma$
guarantees that the slow modes are frozen on the timescale $\Delta t$,
therefore the running average in Eq.~(\ref{runav}) does not mix its
value at time $t$ with the future evolution.

Clearly, as it is, Eq.~(\ref{runav}) is tautological since we usually don't know
$t_\sigma $ unless the slow degrees are defined.
However, using the properties of two-time quantities described
in Sec.~\ref{twotime} and sketched in the left panel of Fig.~\ref{figcorr}, 
the condition~(\ref{condrunav}) can be translated into 
\be
\frac{1}{\Delta t}\int _t ^{t+\Delta t}C(z,t_w)\, dz\simeq C(t+\Delta t,t_w)
\label{condc}
\ee
or equivalently
\be
\frac{1}{\Delta t}\int _{t_w-\Delta t} ^{t_w}C(t,z)\, dz
\simeq C(t,t_w-\Delta t),
\label{condc1}
\ee
for the autocorrelation function
\be
C(t,t_w)=\langle S_i(t)S_i(t_w)\rangle.
\label{autoco}
\ee
This can be understood looking at the schematic plot on the left panel of Fig. \ref{figcorr}. 
Indeed, let us consider the three possible choices of $t$:
i) $t_w<t<t_w+t_\psi$, 
ii) $t_w+t_\psi<t<t_w+t_\sigma $
and iii) $t>t_w+t_\sigma$.  
Starting from ii), in this case 
the correlation function is 
constant and equal to $q$ 
(to be precise this occurs if $t_w+t_\psi<t<t_w+t_\sigma -\Delta t$
but $ t_w+t_\sigma -\Delta t \simeq t_w+t_\sigma $ because of (\ref{condrunav}))
and therefore equations (\ref{condc}) and (\ref{condc1}) are obvious.
On the other hand, considering for simplicity Eq. (\ref{condc}) 
(for Eq. (\ref{condc1}) one can proceed similarly), 
in case i) $C(z,t_w)$ 
is not a constant in the interval $z\in [t_w,t_w+t_\psi]$ but
this integration interval is negligible if the condition 
$\Delta t\gg t_\psi$ in Eq. (\ref{condrunav}) is met. 
Finally, Eq. (\ref{condc}) is true also
in the last case iii), because in this stage $C$ is
decreasing but this occurs on timescales of order $t_\sigma$,
while we have chosen $\Delta t \ll t_\sigma$.  

Letting $t_w=t$ in Eq.~(\ref{condc}), with $t$ chosen as in ii) above, 
one has
\be
C(t+\Delta t,t)\simeq q,
\label{condq}
\ee
which very transparently appears as a direct manifestation of the
choice~(\ref{condrunav}).
The quantity $q$ in this case amounts to the so-called 
Edwards--Anderson order parameter.
In the previous argument and in what follows it is understood that the symbol $\simeq $
is correct in the large-$t_w$ limit, and the corresponding equalities become exact
(i.e. with $\simeq $ replaced by a strict equality $=$) for $t_w\to \infty$.

Let us notice that the modes $\sigma _i,\psi _i$
defined through Eqs.~(\ref{runav},\ref{defpsi}) are statistically
independent. Indeed any cross-correlation between them
vanishes
\be
\langle \sigma _i(t)\psi _i(t_w)\rangle =\langle \psi _i(t)\sigma _i(t_w)\rangle=0,
\label{statind}
\ee
as can be easily proved 
\begin{align}
\langle \sigma _i(t)\psi _i(t_w)\rangle&=
\langle \sigma _i(t)\left [S_i(t_w)-\sigma _i(t_w)\right]\rangle \nonumber \\
&=
\frac{1}{\Delta t}\int _t^{t+\Delta t} C(z,t_w)dz-
\frac{1}{(\Delta t)^2}\int _t^{t+\Delta t}\int _{t_w-\Delta t}^{t_w} C(z,y)dzdy 
\nonumber \\
&\simeq C(t+\Delta t,t_w)-C(t+\Delta t,t_w-\Delta t)\simeq 0,
\end{align}
where we have used Eqs.~(\ref{condc},\ref{condc1}) and the last passage 
can be obtaind by reasoning as below Eq.~(\ref{autoco}) enforcing
the properties of $C$.
The second relation in (\ref{statind}), namely 
$\langle \psi _i(t)\sigma _i(t_w)\rangle=0$, can be proved analogously.

Due to the statistical independence (\ref{statind}), 
the correlation splits into
\be
C(t,t_w)=C_\psi (\tau) +C_\sigma (t,t_w),
\label{split}
\ee
where we have assumed that the $\psi _i$'s are stationary variables.
The schematic behavior of $C_\psi$ and $C_\sigma$ is shown in 
the left panel of Fig.~\ref{figcorr}.

While $\mathcal{O}_\psi$ inherits the properties of the target equilibrium state, 
the aging part $\mathcal{O}_\sigma$ displays new features characteristic of non-equilibrium,
as a remarkable gauge invariance called dynamical scaling.
This can be expressed as the fact that the flow of time can be \emph{fully} taken into account
by changing  the units $\ell $ of measure of  certain dimensional physical quantities. 
Assuming, as it usually (but not necessarily) is, these to be lengths, one has invariant
forms by measuring lengths in unit of the time-dependent unit of measure $\ell (t)$.
For instance, in ferromagnets the equal-time correlation function $G(r,t)=\langle S_i (t)S_j(t)\rangle$
between spins $i,j$ at distance $r$ obeys
\be
G_\sigma (r,t)=\hat g\left [\frac{r}{\ell(t)}\right ],
\label{gscal}
\ee
where $\hat{g}$ is a scaling function. The clear meaning of this is that configurations of
the same system at different times {look the same} on average if we rescale space by 
the characteristic unit measure of lengths, which in this case has the very transparent
meaning of the size of the growing ordered domains of the two symmetry related equilibrium
phases at $T=T_f$. More generally, for a one-time quantity which depends on space, scaling
implies
\be
{\cal O}_\sigma (r,t)=\ell ^{-b}(t)\,\hat o \left [\frac{r}{\ell(t)}\right ],
\label{onesc}
\ee
Eq.~(\ref{gscal}) being a particular case with $b=0$.

In a similar way, for the autocorrelation of a ferromagnet  
one has $C_\sigma (t,t_w)=\hat c\left [\frac{\ell (t)}{\ell(t_w)}\right ]$
which is a particular case (again with $b=0$) of the general scaling
\be
{\cal O}_\sigma (t,t_w)=\ell ^{-b}(t_w)\,\hat o\left [\frac{\ell (t)}{\ell(t_w)}\right ].
\label{twosc}
\ee

In ferromagnetic systems with a scalar order parameter, like those described
by the Ising model, the two classes of degrees $\{ \psi \},\{ \sigma \}$ have a natural interpretation \cite{noipisc}.
After a temperature quench domains of the two equilibrium phases with magnetization
$\pm M$ form, compete and grow. In this case $\sigma _i$ and $\psi _i$ can be thought as
the background magnetization of the domains, which goes from $+M$ to $-M$ passing 
from zero upon moving across a domain's boundary, and the fast thermal flips of spins on top
of this background, respectively. 

Interestingly, a separation of degrees of freedom into two classes with well separated time-scales
can be exhibited in an exactly solvable model  describing a ferromagnetic system with a vectorial
order-parameter with ${\cal N}$ components, in the large-${\cal N}$ limit \cite{noi2002}.
The order parameter $\phi (\vec x,t)$ -- which can be viewed as a coarse-grained
spin -- can be explicitly decomposed as
$\phi (\vec x,t)=\psi (\vec x,t)+\sigma (\vec x,t)$
in an exact analytical way, and the statistical independence~(\ref{statind}) is proved.

The previous considerations show that, in some cases, the splitting of the 
degrees can be achieved. However, in the previous
examples the two kind of modes
are spatially mixed because both components are uniformly spread around the whole system.
In the rest of the paper, instead, we will concentrate on a simple aging system, 
the zeta urn model~\cite{bialas1,bialas2,bialas3,godreche1,godreche2,godreche3,godreche4} which, besides the advantage of being analytically
tractable, has the remarkable property that $\{ \psi \}$ and $\{ \sigma \}$ are segregated
in different parts of the system. This allows to concretely visualize them and offers a more
intuitive interpretation.

\section{The zeta urn model}\label{urnmod}

The zeta urn model is defined by a collection $r_i \in \{0,1,\dots, M\}$ 
of integer variables
defined on a network of $N$ nodes $i$.
Usually the nodes are identified as boxes or urns inside which $r_i$
marbles, whose number $M= \sum_{i=1}^N r_i$ is fixed and conserved, 
can be accommodated.  
A cost function, or Hamiltonian, is introduced as the sum of independent contributions from
all the nodes
\be
{\cal H}(\{r_i\})=\sum _{i=1}^N E_i,
\label{ham}
\ee
where 
\be
E_i=\ln (r_i+1)
\ee
is the (energy) cost to accommodate $r_i$ marbles inside the $i$-th box.
In a statistical mechanical approach one introduces a canonical ensemble
at the inverse temperature $\beta $ and configurations $\{r_i\}$ are
weighted by the Boltzmann measure
\be
{\cal P}(\{r_i\})=Z^{-1}(\beta ,N,M)e^{-\beta {\cal H}(\{r_i\})}
\ee
where
\be
Z(\beta,N,M)=\sum _{\{r_i\}}e^{-\beta {\cal H}(\{r_i\})}
\ee
is the partition function.
An exact computation shows the existence of a density-dependent critical temperature
$T_c=\beta _c^{-1}$. Above such temperature  the probability $P(r)$ of any of the 
identical urns to contain
$r$ marbles decays exponentially with $r$, while 
\be
P(r)\sim (r+1)^{-\beta },
\label{fcrit}
\ee
with $\beta =\beta _c$, at
the critical point. In the low-temperature phase a macroscopic condensate
appears: the occupation probability $P(r)$ is still given by Eq.~(\ref{fcrit}),
with the current value of $\beta $, 
for all the boxes except one, which contains an extensive number of marbles.
 
The equilibrium measure is invariant under a dynamics that respects detailed balance.
In the following we will consider a kinetic rule where the number of particles
is conserved and the transition rate $w(r_i,r_j)$ to move a particle
from a non-empty urn $i$ to another $j$, chosen randomly in the network and containing 
respectively $r_i$ and $r_j$ particles before the move, 
is given by the Metropolis expression
\be
w(r_i,r_j)=\min \left [1,\exp (-\beta \Delta E)\right ],
\label{trrate}
\ee
where
\be
\Delta E=\ln (r_i)-\ln (r_i+1)+\ln (r_j+2)-\ln (r_j+1)
\ee
is the variation of the energy due to the move.
Notice that the present kinetic rule has a mean field nature, in the sense that
particles can be transferred between any couple of nodes.

The dynamics of the present model was solved exactly in different conditions
in  Ref.~\cite{godreche4} where
many quantities were computed. Here, since we are interested in separating
degrees of freedom with different properties, we study it numerically. 

\section{Numerical study of the model} \label{numerical}

We performed a series of simulations of a system of $N=500$ urns, which
are initially prepared in an infinite temperature equilibrium condition
with particles distributed according to
\be
P(r)=\frac{\rho ^r}{(\rho +1)^{r+1}},
\ee
where $\rho =M/N$ is the density which, in the following, we will set to $\rho =2$.
From this initial condition we quench the system to the final inverse temperature
$\beta _f=(k_BT_f)^{-1}=6>\beta _c$, meaning that we evolve the dynamics 
with this value of $\beta $
in the transition rates (\ref{trrate}). Observables are averaged over a 
non-equilibrium ensemble of $3\times10^5$ realizations of the process with different
initial conditions and thermal histories.

The target equilibrium state with a macroscopic number of marbles 
condensed in a single urn is approached   
by a dynamical process that can be roughly divided into three stages.
In an early stage, for $0<t\ll N$ (where $t$ is the time elapsed after the
quench, measured in units of Monte Carlo sweeps), a segregation occurs between urns 
with a low occupation
number, that we will denote as the {\it gas} phase in the following,
and a {\it liquid} phase containing boxes occupied by many marbles.
For larger times $N\ll t\ll N^2$ a scaling regime sets in during which
the gaseous regions remain on average stationary while urns of the
liquid phase compete and, as an effect of this, their number
diminishes. Eventually, for $t\gg N^2$ a finite-size effect 
sets in, the symmetry between the 
nodes is spontaneously broken, a finite fraction of marbles is
stored in a prevailing urn and the rest of them is dispersed
among the remaining $N-1$ boxes. 

\subsection{Occupation probability} \label{occu}

\begin{figure}
    \centering
\rotatebox{0}{\resizebox{.8\textwidth}{!}{\includegraphics{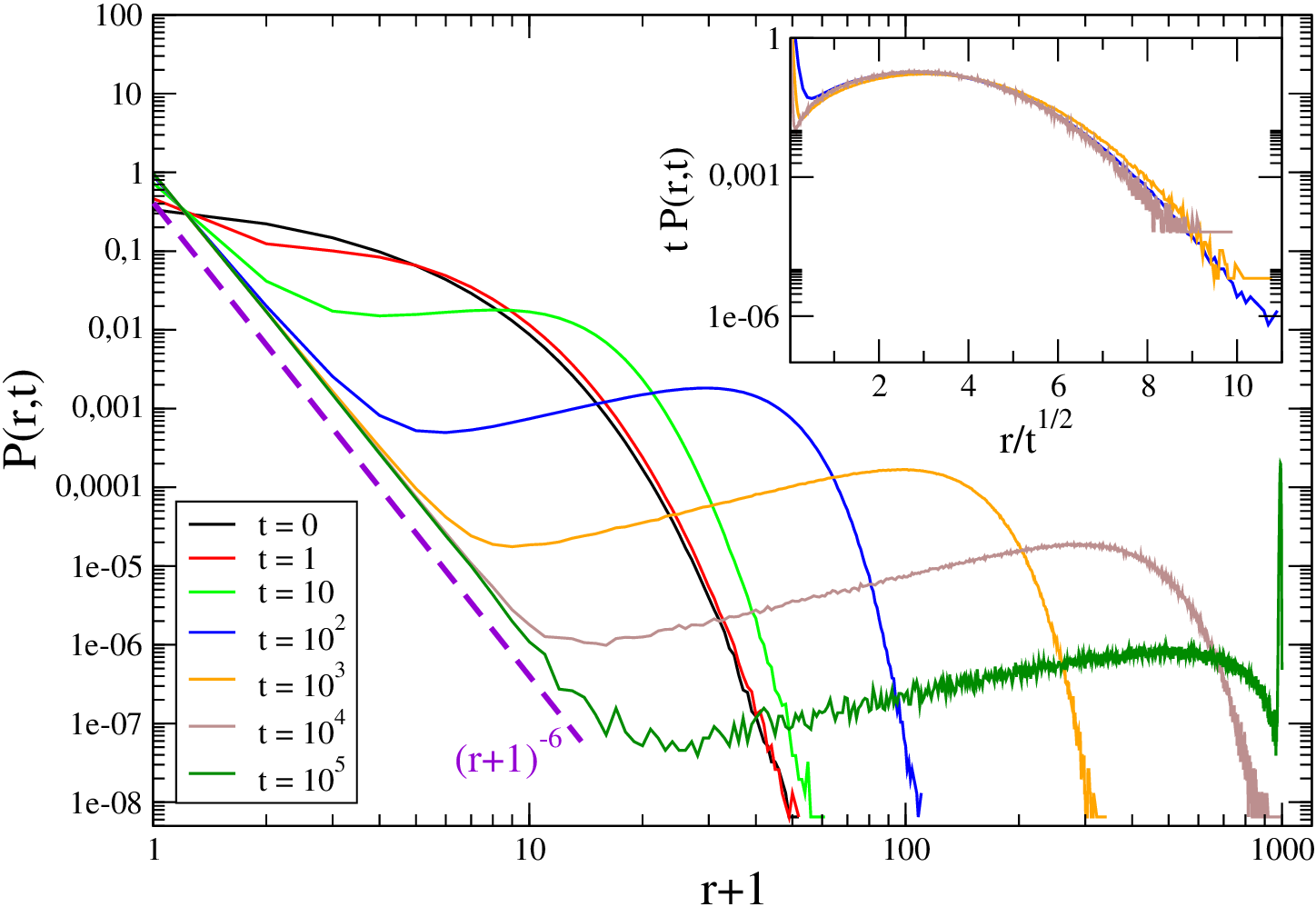}}}
\caption{The occupation probability $P(r,t)$ is plotted against $r+1$
on a log-log plot for $\beta _f=6$ and $\rho =2$. 
The dashed violet line is the law $(r+1)^{-6}$,
namely the equilibrium occupation probability
(\ref{fcrit}) of the target equilibrium state.
In the inset $t\,P(r,t)$ is plotted against $r/t^{1/2}$.}
\label{fighist}
\end{figure}

This whole pattern of behavior is mirrored by the evolution of
the occupation probability $P(r,t)$, 
as it is shown in Fig.~\ref{fighist}.
From the initial distribution, $P(r,t)$ starts to progressively
form a hollow at an intermediate value $r_{min}$ of $r$, signalling that the
segregation process is started, and around $t\simeq 10^2$
a secondary peak is developed around $r=r_{max}=10$. From this time onward 
the peak moves progressively to larger values of $r$ and becomes
more pronounced as compared to the hollow minimum.
We will say that urns with $r<r_{min}$ are in the gaseous phase and the remaining
ones in the liquid one.

Around $t\simeq t_{sc}\simeq 1.5\cdot 10^2$ the second regime is entered,
where the form of the different
curves do not seem to change except for a rescaling of both axis.

Eventually, around $t=10^5$ a sharp peak located at the upper right of the
spectrum starts growing. This is the hallmark of the final stage of
the process where the finite size is felt and all the condensed
marbles fill a single urn.

In the following we will concentrate on the second regime where the scaling
properties are observed. Let us stress that, if the thermodynamic limit
$N\to \infty$ (with fixed $\rho$) is taken from the onset, 
this stage becomes truly asymptotic. 
As discussed in Secs.~\ref{intro}, \ref{twotime}, \ref{statag}, this is the regime where the two classes of
degrees $\{ \psi \}$ and $\{ \sigma \}$ coexist. 

We show now that these two sets
can be consistently associated to the gaseous and the liquid urns.
In order to do that, we observe that -- as it can be seen in 
Fig. \ref{fighist} -- for $r<r_{min}$
the curves are time-independent and all collapse on the equilibrium power-law
distribution (\ref{fcrit}). Therefore urns with $r\ll r_{min}$ are already in
an equilibrium state and this is the characteristic property of the $\{ \psi \}$ component.
Conversely -- as it is shown in the inset of Fig. \ref{fighist} -- 
by plotting $t\, P(r,t)$ vs $r/t^{1/2}$, for $r\gg r_{min}$ the curves at different times
collapse on a single master curve. This means that the form
\be
P_\sigma(r,t)=\ell^{-2}(t)\,\hat p\left (\frac{r}{\ell(t)}\right )
\label{scalhist}
\ee
is obeyed, where \cite{godreche1}
\be
\ell (t)=t^{1/2}
\label{ell}
\ee
is a time-dependent characteristic marble number and $\hat p$ is a scaling function.

Equation (\ref{scalhist}) is a typical scaling property, 
formally identical to Eq.~(\ref{onesc}), with the difference that $r$ is not
a space coordinate but, say, the height of the stack of marbles in an urn and 
$\ell (t)$ the unit to measure it. 
This suggests the quite natural identification of the highly populated urns  
with $r>r_{min}$ as $\{ \sigma \}$ (hence the subscript in $P_\sigma$)\footnote{More precisely, 
one should say that the highly populated urns contain also,
besides the condensate $\{\sigma \}$, a fraction of the gas
phase. The latter is represented by the few marbles hopping
from/to the scarcely populated urns. However, for sufficiently
low temperatures this gaseous fraction is very small (less then
one ball in a condensed urn, in our simulations),
so we can neglect it and identify the highly populated urns as
the condensed phase.}. 
Notice that the property (\ref{scalhist}) implies
that the number $n_\sigma(t) \simeq \sum _{r=r_{min}}^N P(r,t)$ of liquid urns 
decreases as 
\be \label{liqzero}
	n_\sigma(t) \sim \ell (t)^{-1}\,,
\ee 
as it is expected for an aging component according to Eq. (\ref{nsig0}). 
Since the number of marbles in the condensate is roughly constant during the scaling 
regime, the shrinking of $n_\sigma(t)$ entails a correspondent growth of the occupation number 
of a liquid urn proportional to $\ell(t)$. This justifies the physical interpretation of $\ell(t)$
as the natural unit of measure of $r$.

The scaling (\ref{scalhist}) implies also that $\lim _{t\to \infty} P_\sigma(r,t)=0$, that is
the property (\ref{eqquant}), hence $P(r,t)$ is an equilibrating quantity. 
However quantities like the fluctuation 
$\langle r^2 \rangle _\sigma - \langle r\rangle _\sigma ^2$ 
of the number of liquid marbles (where $\langle r\rangle _\sigma= \sum_\sigma r\,P(r,t)$ and
similarly for $\langle r^2\rangle _\sigma $) are not equilibrating,
as it can easily be checked using Eq. (\ref{scalhist}).

\subsection{Energy}

\begin{figure}
    \centering
{\resizebox{.8\textwidth}{!}{\includegraphics{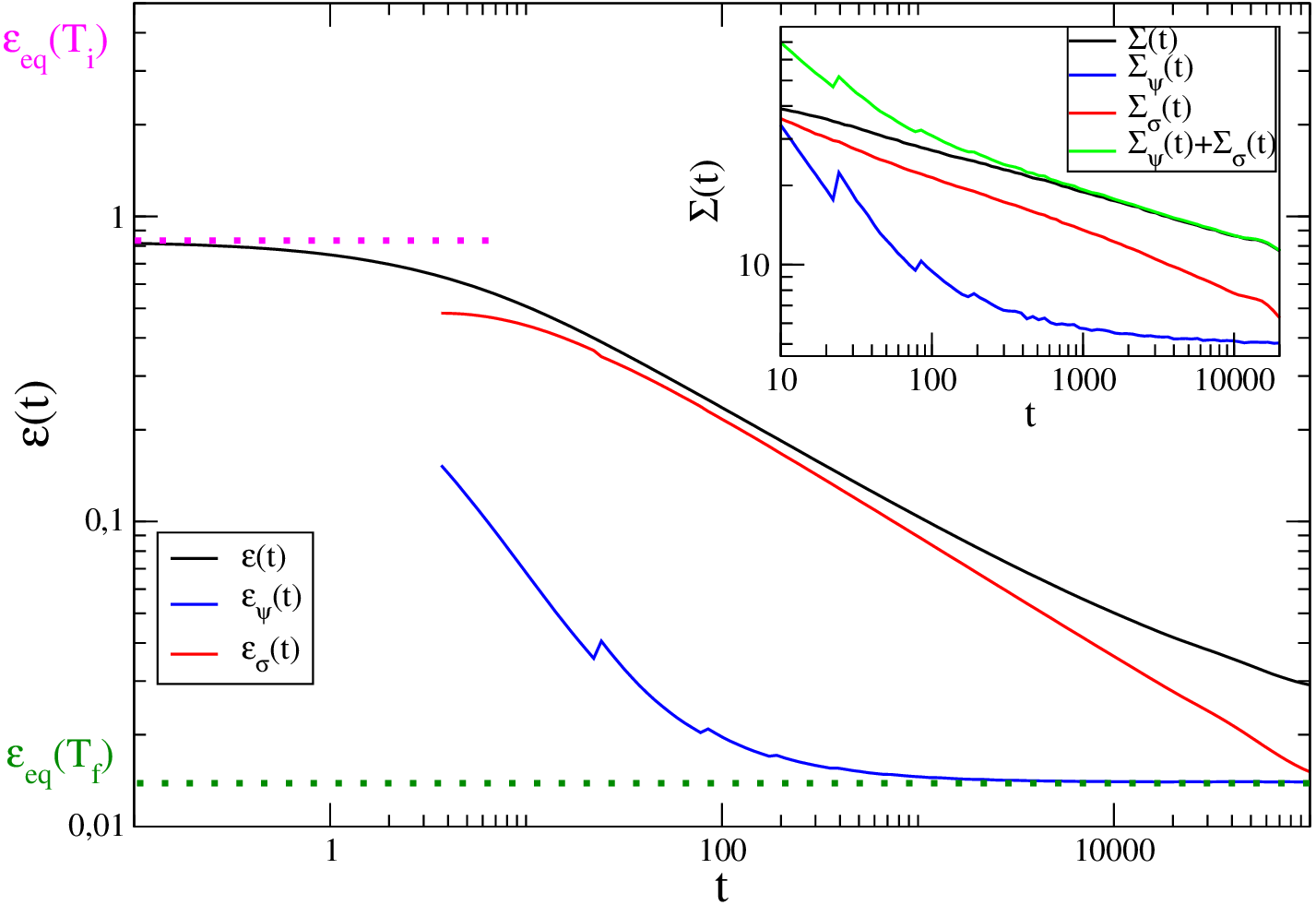}}}
\caption{The quantities $\varepsilon(t)$, $\varepsilon_\psi (t)$ and $\varepsilon_\sigma (t)$ are plotted against
$t$ in a double logarithmic plot. The dashed lines are the analytically known equilibrium values
$\varepsilon_{eq}(T_i)$ and $\varepsilon_{eq}(T_f)$ at the initial and final temperatures, respectively. In the inset the variances $\Sigma (t)$, $\Sigma _\psi (t)$, 
$\Sigma _\sigma (t)$ are shown together with the sum of the last two.}
\label{figenergy}
\end{figure}

As a different example of a one-time quantity we have considered the energy density
$\varepsilon(t)=E(t)/N=\langle {\cal H}\rangle/N$ of the system, where the Hamiltonian is given in Eq.~(\ref{ham}).
Following the decomposition (\ref{additive}) we write $\varepsilon(t)=\varepsilon_\psi(t)+\varepsilon_\sigma(t)$ and,
using the non-interacting character (\ref{ham}) of the model, we define
$\varepsilon_\psi =\sum _{i\in \{\psi\}}E_i/N$ and $\varepsilon_\sigma =\sum _{i\in \{\sigma\}}E_i/N$
(the two sets $\{\psi \}$ and $\{\sigma \}$ are identified as discussed above).
These quantities are shown in Fig. \ref{figenergy}. $\varepsilon_\psi (t)$ converges to 
the final equilibrium value $\varepsilon_{eq}(T_f)=-\zeta'(\beta)/\zeta(\beta)$
in a time of order $t\simeq 10^3$. 
$\varepsilon_\sigma (t)$ decays to zero proportionally to
$(\ln t)/\sqrt{t}$, as can be checked using the scaling form (\ref{scalhist}), showing that
the internal energy is an equilibrating quantity and that, for large time, the contribution
of $\varepsilon_\sigma $ can be neglected in $\varepsilon$. 

Notice that the property $\lim _{t\to \infty}\varepsilon_\psi (t)=\varepsilon_{eq}(T_f)$ is an independent
confirmation of the reliability of the method proposed above to separate fast and
slow degrees.

We have also computed the variance 
$\Sigma (t)=\langle {\cal H}^2\rangle - \langle {\cal H}\rangle ^2$
of the energy and of its gaseous and liquid parts,
$\Sigma _\psi (t)=\langle {\cal H}^2\rangle _\psi - \langle {\cal H}\rangle _\psi ^2$ and
$\Sigma _\sigma(t)=\langle {\cal H}^2\rangle _\sigma - \langle {\cal H}\rangle _\sigma ^2$,
respectively. These quantities are shown in the inset of 
Fig. \ref{figenergy}. Here one sees that for sufficiently long times
($t \gtrsim 500$), when the separation of degrees becomes effective,
$\Sigma (t)=\Sigma _\psi (t)+\Sigma _\sigma (t)$. This is a nice
indication that the $\psi$'s and the $\sigma$'s are statistically independent
variables. 

\subsection{Autocorrelation function}

As an example of two-time quantity we consider the autocorrelation function.
The number-number correlation 
\be
\widetilde C(t,t_w)=\langle r_i (t)r_i(t_w)\rangle -\rho ^2,
\label{cgod}
\ee
which does not depend on $i$ due to homogeneity,
was analytically worked out in \cite{godreche4}. However this quantity is not 
well suited to the purposes of our paper because it is entirely dominated
by the contribution of the highly populated urns forming the condensate.
For this reason, instead of 
$\widetilde C$, we will consider
\be
C(t,t_w)=\langle \delta r_i(t)\delta r_i(t_w)\rangle,
\label{newcor}
\ee
 where $\delta r_i(t)=r_i(t+1)-r_i(t)$, is the variation of the number of marbles in the
 $i$-th urn. 
This quantity is related to $\widetilde C$ by
\be
C(t,t_w)=\widetilde C(t+1,t_w +1)-\widetilde C(t+1,t_w)-\widetilde C(t,t_w +1)+
\widetilde C(t,t_w).
\label{tildevsno}
\ee 
The definition (\ref{newcor}) is inspired by the autocorrelation encountered 
in the study of ferromagnetic systems.
 Typical values taken by the $\delta r$'s are $\delta r_i(t)=0,\pm 1$
(any value between $-M$ and $M$ is in principle possible, but with a very small probability) for all the urns,
 and this makes the difference with the definition (\ref{cgod}) where the largest contribution 
 is provided by large values of $r_i$ from the urns in the liquid phase. 

Let us now discuss the splitting (\ref{split}) of this autocorrelation. 
The gaseous part $C_\psi$ has very different features with respect to the liquid one 
$C_\sigma$.
If an urn is in the gas phase, the number of marbles contained in it fluctuates rapidly
around the average gas occupation number $\rho _c=\zeta(\beta -1)/\zeta(\beta)-1\simeq 0.019$ (where $\zeta$ is the Riemann's zeta-function), hence $C_\psi $ is expected 
to converge quite rapidly (over a microscopic 
typical time of the equilibrium state) to zero. This decay occurs from the negative side,
because if by chance an urn with an average number of balls receives an extra marble,
it will most probably release it in the future in order to keep its occupation number 
close to the mean. Furthermore, on general grounds, $C_\psi$ is expected to depend only
on the time difference $\tau$.

Conversely, in the effort to build the condensate, one urn in the liquid phase
have the tendency to increase its number of marbles. This occurs up to a long time
(of order $t_w$) when the competition with another prevailing urn will become 
effective and it will be progressively depleted.
As a consequence, $C_\sigma$ is expected to show a pronounced maximum
at a time monotonously increasing with $t_w$ and then to
decrease to zero.

\begin{figure}
    \centering
{\resizebox{.8\textwidth}{!}{\includegraphics{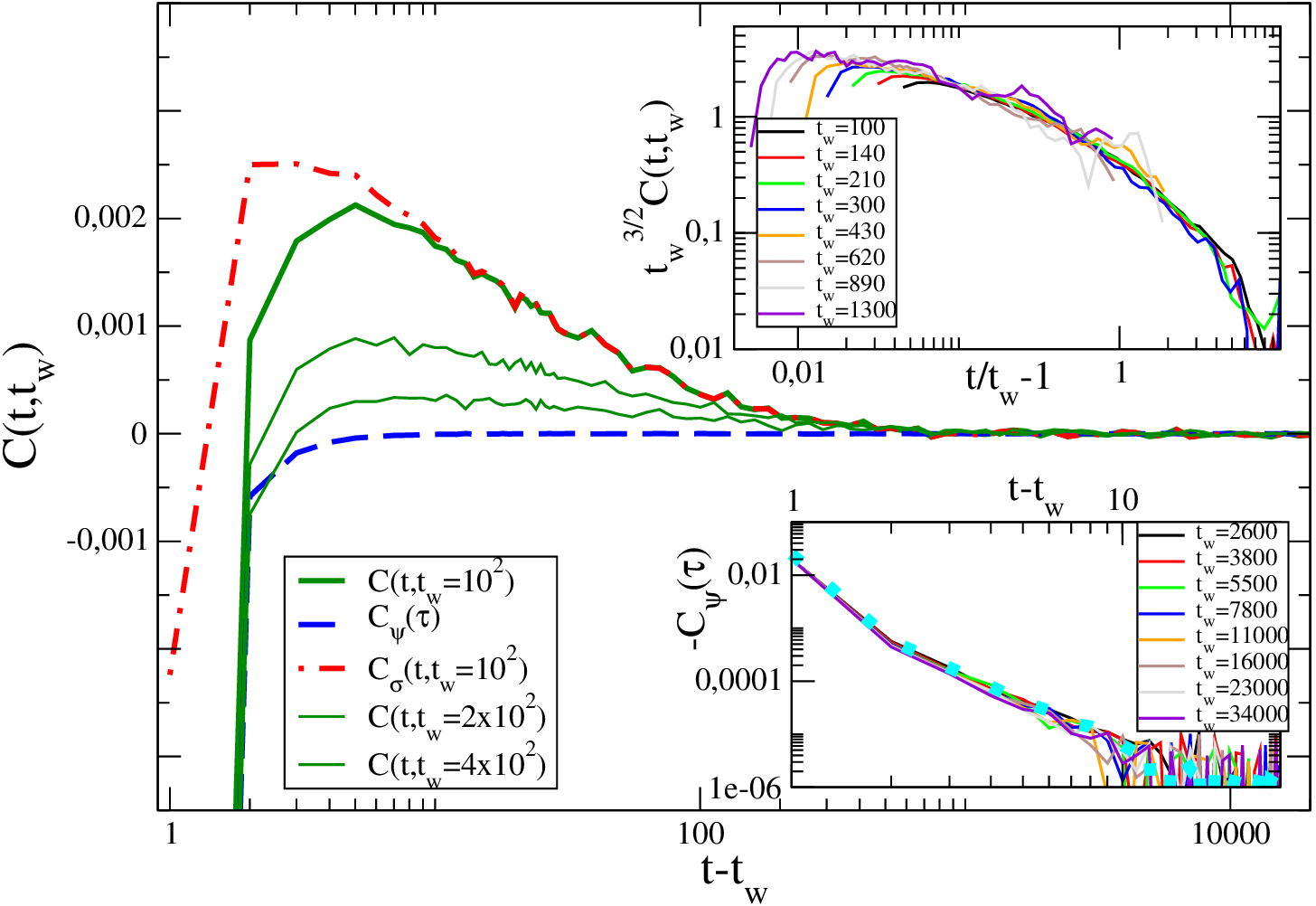}}}
\caption{The autocorrelation $C(t,t_w)$ (continuous green curves) is plotted against $t-t_w$ 
on a semi-log scale, for three different waiting times (lower curves correspond to 
larger $t_w$). The long-dashed blue curve is $C_\psi (\tau)$ and the dot-dashed 
red one is $C_\sigma (t,t_w)$.
In the upper inset the data collapse of the same curves of the left panel is obtained
by plotting $t_w^{3/2}C(t,t_w)$ against $(t-t_w)/t_w$.
In the lower inset $-C_\psi (t,t_w)$ is plotted for different choices of $t_w$ (see key).
The cyan dots denote $-C_{eq}(\tau)$.}
\label{figcorurns}
\end{figure}

All these features are clearly displayed in the main part of Fig. \ref{figcorurns}, where 
the autocorrelation is plotted against $t-t_w$ for different values of $t_w$. 
Let us start by considering the solid green lines, which correspond
to the whole correlation $C(t,t_w)$.

The curves relative to different values of $t_w$ collapse for short times, up to $t-t_w\simeq 2$.
In this regime $C(t,t_w)$ is negative. What we are observing in this region is clearly the 
evolution of the fast part $C_\psi (\tau)$. 

From $t-t_w\simeq 2$ onward the curves
grow to a positive maximum, which is delayed and suppressed as $t_w$ is increased,
and then decrease to zero.
This is the behavior expected qualitatively for $C_\sigma$.
Furthermore, in this regime the curves for different $t_w$ can be collapsed by plotting 
$t_w^{3/2}C(t,t_w)$, against $t/t_w$, as it can be seen in the inset of Fig.~\ref{figcorurns}.
This means that the scaling form
\be
C(t,t_w)\simeq \ell^{-3}(t_w)\,\hat c\left (\frac{t}{t_w}\right )
\label{scalpart}
\ee
is obeyed, which is of the expected type (\ref{twosc}), with the behavior (\ref{ell}) of 
$\ell $ previously found and $b=3$, yet another consequence of the decrease
(\ref{liqzero}) of urns in the liquid phase. 

Let us mention that the form (\ref{scalpart}) is in agreement with the
known behavior 
\be
\widetilde C(t,t_w)=(\rho -\rho_c)\,t_w^{1/2}\,\Phi(t/t_w),
\ee
(where $\Phi (x)$ is a scaling function)
of $\widetilde C(t,t_w)$ in the aging regime.
Indeed, by plugging this expression into Eq. (\ref{tildevsno}) and expanding
for large $t_w$
the various correlations on the r.h.s. around the value $(t,t_w)$ of their arguments,
one finds Eq. (\ref{scalpart}) with 
$\hat c=-\frac{\rho -\rho _c}{2}\,\Phi '(t/t_w)$,
where $\Phi '(x)=d\Phi(x)/dx$.

In order to check the decomposition (\ref{split}) we have separated the urns into the
gaseous and the liquid phase as discussed previously, and we have separately computed
$C_\psi$ and $C_\sigma$ by correlating among them degrees that pertain to one or
the other set both at $t_w$ and $t$. 
These quantities are also plotted in Fig.~\ref{figcorurns}, where it is seen that they 
reproduce the whole quantity $C$ in the short-time ($C_\psi$) and in the long-time
regime ($C_\sigma $), respectively. For $C_\sigma$ we have also checked that the scaling
(\ref{scalpart}) is obeyed.

In order to better compare the behavior of 
the autocorrelation in the present urn model with that of a ferromagnet we have
represented it pictorially in the right panel of Fig. \ref{figcorr} using 
the same line styles and colors in the two cases. The analysis of  the two panels
shows that a similar scenario occurs, apart of course from the different
functional form of the various correlations.

According to the general paradigm discussed 
in Sections \ref{intro}, \ref{twotime}, \ref{statag}, 
$C_\psi (\tau)$ should correspond to the correlation 
function of the target equilibrium state.
In the lower inset of Fig.~\ref{figcorurns} we plot this quantity computed in the
aging regime by separating the degrees of freedom for different choices of $t_w$.  
The data do not show any systematic dependence
on $t_w$, confirming that our definition of the set $\{\psi\}$ has the expected
stationary property.
In order to compare $C_\psi$ with the equilibrium autocorrelation we have 
computed the latter numerically: An efficient way to
equilibrate the system at $T<T_c$ is to break the urn symmetry by starting from an
initial configuration with all the marbles in one urn. This avoids the aging regime
caused by the competition among urns and allows the system to quickly reach the
equilibrium state. The quantity $C_{eq}(\tau)$ measured in this way is shown in 
the lower inset of Fig.~\ref{figcorurns} with a heavy-dashed cyan curve.
As expected, $C_\psi (\tau)$ and $C_{eq}(\tau)$ coincide within numerical errors.
The equilibrium autocorrelation could also be computed analytically starting from the
known \cite{godreche4} expression of $\widetilde C$
\be
\widetilde C_{eq}(\tau)=A_{eq}(\beta_f) \tau^{-(\beta -3)/2},
\ee
where $A_{eq}(\beta)$ is a temperature-dependent constant.
From this it is trivial to derive the expression
\be
C_{eq}(\tau)\simeq -\frac{(\beta -1)(\beta -3)}{4} A_{eq} (\beta _f)\tau ^{-(\beta +1)/2}
\label{analceq}
\ee
for $C_{eq}(\tau)$. We have checked that our numerical determination shown in the
lower inset of Fig.~\ref{figcorurns}
coincides within errors with Eq.~(\ref{analceq}).

Summarizing, our numerical data for the zeta urn model fit nicely in the general framework
of slow relaxation discussed in Sections \ref{intro}, \ref{twotime}, \ref{statag} if the
degrees of freedom of the system -- the occupation numbers $r_i$ -- are divided into two sets $\{\psi\}$ and $\{\sigma\}$ 
according to the recipe described in Section \ref{occu}. 

Before closing this section, we briefly comment on the robustness of  
the overall
picture with respect to different definitions -- but same in spirit -- of the 
fast and slow degrees. We tried to address this issue by changing the definition of 
$\psi$'s and $\sigma$'s and, next to the technique of Section \ref{occu}, we have
separated slow and fast degrees using the running average procedure 
Eqs.~(\ref{runav},\ref{defpsi}) of Section \ref{statag}. More precisely, 
considering the autocorrelation function $C$
we have defined the slow part as 
\be
C_\sigma (t,t_w)=\langle \Delta r_i (t)\Delta r_i (t_w)\rangle
\label{otherslow}
\ee
where $\Delta r_i(t)=(\Delta t)^{-1}[r_i(t+\Delta t)-r_i(t)]$ is the variation of 
number of marbles (normalized by $\Delta t$)
in the $i$-th urn in a time $\Delta t$ such that -- as discussed in Section \ref{statag} --
$C_\psi (\tau=\Delta t)\simeq 0$. Inspection of Fig.~\ref{figcorurns} suggests
$\Delta t \sim 10\div50$ as reasonable values. By computing the quantity of 
Eq.~(\ref{otherslow}) we found that, for sufficiently
large values of $t-t_w$ the curve obtained superimposes on 
the determination of $C_\sigma $ 
obtained previously with the technique of Section \ref{occu}. 
This results confirms the robustness of the degree-separation scenario.

\section{Conclusions} \label{concl}

In this paper we have discussed the paradigm of a simple class of aging
systems where the dynamical properties can be accounted for by the 
coexistence of slow and fast degrees with different scaling properties.

We have studied -- in particular -- the kinetics of the zeta urn model
following a quench from high temperature to a final one
below the condensation temperature.
This model is a useful playground where fast and slow degrees have an
intuitive interpretation in terms of the occupation numbers of different
urns, allowing a distinction between a gaseous and a liquid phase,
corresponding to the two sets $\{\psi\}$ and $\{\sigma \}$ of
fast and slow modes. 

We have considered the behavior of one-time quantities,
as the energy, and of two-time ones, specifically the autocorrelation function.
Due to the progressive reduction of the {\it measure} of the aging set $\{\sigma \}$,
one-time quantities rapidly attain the characteristic value of the final
equilibrium state. Non-equilibrium features are better observed in two-time
quantities where the gaseous and the liquid component can be evidenced
by increasing the two times along particular directions in the $(t,t_w)$ plane,
namely focusing on the short-time or on the aging regime, very similarly
to what is known for ferromagnetic or glassy spin systems.

The concept of an \emph{effective temperature} is deeply related to the 
topic considered in the present Article. This quantity, 
which  was originally introduced 
in the context of mean-field glass models \cite{cuglia1,cuglia2,cuglia3,cuglia4,cuglia5}, can be obtained from the relation
between an autocorrelation function and the correspondent response function.  
For systems -- like the one considered here -- where two sets of degrees can be
identified, one would  expect two different \emph{temperatures} associated to them.
Accordingly, while the set $\{\psi \}$ is equilibrated at the bath temperature,
a different \emph{effective} temperature would characterize the 
slow part $\{ \sigma \}$. An effective temperature for the zeta urn model based on the 
correlation (\ref{cgod}) has already been investigated \cite{godreche4}: The different insight 
that could be provided by starting from our new definition (\ref{newcor}) instead
remains an open issue for future research.

\section*{Acknowledgments}
F.C. acknowledges financial support from MIUR PRIN 2010HXAW77\_005. 
G.G. acknowledges the support of MIUR (project PRIN 2012NNRKAF).
Part of this research took place at the Galileo Galilei Institute for Theoretical 
Physics in Arcetri, during the INFN-funded summer 2014 workshop 
\emph{Advances in Nonequilibrium Statistical Mechanics}.



\begin{thebibliography}{99}

\bibitem{slow1} Bouchaud JP, Cugliandolo LF, Kurchan J, Mezard M. Out of equilibrium dynamics in spin-glasses and other glassy systems, in Young AP (Ed.), Spin Glasses and Random Fields, World Scientific, Singapore, 1997.
\bibitem{slow2} Biroli G. A crash course on aging, 2005 (arXiv:cond-mat/0504681).
\bibitem{slow3} Cugliandolo LF. Dynamics of glassy systems, in Barrat J-L, Dalibard J, Kurchan J,
Feigel’man MV (Eds.), Slow Relaxation and Nonequilibrium Dynamics in Condensed Matter, Les Houches - {\'E}cole d'{\'E}t{\'e} de Physique
Theorique, Vol. 77, Springer-Verlag, 2003 (arXiv:cond-mat/0210312).
\bibitem{slow4} Zannetti M. Aging in domain growth, in Puri S, Wadhawan V (Eds.), Kinetics of Phase Transitions, 
CRC Press, 2009 (arXiv:1412.4670).
\bibitem{slow5} Corberi F, Cugliandolo LF, Yoshino H. Growing length scales in aging systems, in Berthier L, Biroli G, Bouchaud J-P,  Cipelletti L, van Saarloos W (Eds.), Dynamical heterogeneities in glasses, colloids, and granular media, Oxford University Press, 2010 (arXiv:1010.0149).
\bibitem{slow6} Henkel M, Pleimling M. Non-equilibrium phase transitions. Volume 2: Ageing and dynamical scaling far from equilibrium, Springer, Dordrecht, 2010.
\bibitem{MVZ} Mazenko GF, Valls OT, Zannetti M. Field theory for growth kinetics. Phys. Rev. B 1988; 38(1): 520--542.
\bibitem{FV} Franz S, Virasoro MA. J. Phys. A: Math. Gen. 2000; 33(5): 891--905.
\bibitem{noi2002} Corberi F, Lippiello E, Zannetti M. Slow relaxation in the large-$N$ model for phase ordering. Phys. Rev. E 2002; 65: 046136.
\bibitem{bialas1} Bialas P, Burda Z, Johnston D. Condensation in the Backgammon model. Nucl. Phys. B 1997; 493: 505--516. 
\bibitem{bialas2} Bialas P, Burda Z, Johnston D. Phase diagram of the mean field model of simplicial gravity. Nucl Phys B 1999; 542: 413--424. 
\bibitem{bialas3} Bialas P, Bogacz L, Burda Z, Johnston D. Finite size scaling of the balls in boxes model. Nucl Phys. B 2000; 575: 599--612.
\bibitem{noirf1} Corberi F, de Candia A, Lippiello E, Zannetti M. Off-equilibrium response function in the one-dimensional random-field Ising model. Phys. Rev. E 2002; 65: 046114.
\bibitem{noirf2} Corberi F, Lippiello E, Zannetti M. On the connection between off-equilibrium response and statics in non disordered coarsening systems. Eur. Phys. J. B 2001; 24: 359--376. 
\bibitem{noirf3} Corberi F, Lippiello E, Zannetti M. Interface fluctuations, bulk fluctuations, and dimensionality in the off-equilibrium response of coarsening systems. Phys. Rev. E 2001; 63: 061506.
\bibitem{noirf4} Corberi F, Lippiello E, Zannetti M. Fluctuation dissipation relations far from equilibrium. J. Stat. Mech 2007; P07002.
\bibitem{noirf5} Corberi F, Castellano C, Lippiello E, Zannetti M. Generic features of the fluctuation dissipation relation in coarsening systems. Phys. Rev. E 2004; 70: 017103.
\bibitem{noipisc} Corberi F, Gonnella G, Piscitelli A. Heat exchanges in coarsening systems. J. Stat. Mech. 2011; P10022.
\bibitem{godreche1} Drouffe J-M, Godr\`eche C, Camia F. A simple stochastic model for the dynamics of condensation. J. Phys. A: Math. Gen. 1998; 31(1): L19--L25.
\bibitem{godreche2} Godr\`eche C. From urn models to zero-range processes: Statics and dynamics, in Henkel M, Pleimling M, Sanctuary R (Eds.), Ageing and the Glass Transition, Lect. Notes Phys. 716, Springer 2007. 
\bibitem{godreche3} Godr\`eche C, Luck JM. Nonequilibrium dynamics of urn models. J. Phys.: Condens. Matter 2002; 14(7): 1601--1615. 
\bibitem{godreche4} Godr\`eche C, Luck JM. Nonequilibrium dynamics of the zeta urn model. Eur. Phys. J B 2001; 23(4): 473--486.
\bibitem{cuglia1} Cugliandolo LF, Kurchan J. Analytical solution of the off-equilibrium dynamics of a long-range spin-glass model. Phys. Rev. Lett. 1993; 71(1): 173--176.
\bibitem{cuglia2} Cugliandolo LF, Kurchan J. On the out-of-equilibrium relaxation of the Sherrington--Kirkpatrick model. J. Phys. A: Math. Gen. 1994; 27(17): 5749--5772. 
\bibitem{cuglia3} Crisanti A, Ritort F. Violation of the fluctuation--dissipation theorem in glassy systems: basic notions and the numerical evidence. J. Phys. A: Math. Gen. 2003; 36(21): R181--R290. 
\bibitem{cuglia4} Cugliandolo LF, Kurchan J, Peliti L. Energy flow, partial equilibration, and effective temperatures in systems with slow dynamics. Phys Rev. E 1997; 55(4): 3898--3914.
\bibitem{cuglia5} Corberi F, Lippiello E, Zannetti M. The effective temperature in the quenching of coarsening systems to and to below $T_c$. J. Stat. Mech. 2004; P12007.

\end{thebibliography}
\end{document}